\documentclass[preprint,showpacs,preprintnumbers,amsmath,amssymb]{revtex4}

\usepackage{graphicx}
\usepackage{dcolumn}
\usepackage{bm}
\usepackage{latexsym}
\usepackage{amsfonts}
\usepackage{amssymb}
\usepackage{amsmath}
\usepackage[usenames]{color}

\begin{document}
\preprint{KUNS-2360}

\title{ Pathologies in Lovelock AdS Black Branes and AdS/CFT }

\author{Tomohiro Takahashi}
\author{Jiro Soda}

\affiliation{Department of Physics,  Kyoto University, Kyoto, 606-8502, Japan
}

\date{\today}

\begin{abstract}
We study the pathologies in AdS black branes in Lovelock theory. 
More precisely, we examine the conditions that AdS 
black branes have the naked singularity, 
the ghost instability and the dynamical instability. 
From the point of view of the AdS/CFT correspondence,
 the pathologies in AdS black branes indicate the pathologies in the corresponding CFT.
Hence, we need to be careful when we apply AdS/CFT in Lovelock theory
to various phenomena such as the
shear viscosity to entropy ratio in strongly coupled quantum filed theory. 
\end{abstract}

\maketitle

\section{Introduction}

It is well known that AdS black branes play a central role in the application
of the AdS/CFT correspondence to various phenomena such as condensed matter physics and fluid dynamics~\cite{Hartnoll:2009sz,Herzog:2009xv,Horowitz:2010gk}. 
Remarkably, the AdS/CFT correspondence holds in any dimensions. 
In higher dimensions, however, a natural theory of gravity is not general relativity
 but Lovelock theory~\cite{Lovelock:1971yv,Charmousis:2008kc}. 
Thus, it is natural to consider the AdS/CFT correspondence
 in the context of Lovelock theory. 

The AdS/CFT correspondence in Lovelock theory 
has been already discussed in the context of the shear viscosity to entropy ratio.
It is conjectured that the shear viscosity to entropy ratio $\eta/s$ 
is larger than $1/4\pi$, which is called as the KSS bound~\cite{Kovtun:2004de}. 
Recently, in the case that the dual gravitational theory is Lovelock theory, 
this ratio has been calculated as $\eta/s=(1-2\alpha_2)/4\pi$, where $\alpha_2$ is the appropriately normalized second order Lovelock coefficient~\cite{Shu:2009ax}.  
It seems that the KSS bound is violated for a positive $\alpha_2$. 
However, when we take into account the pathologies in AdS black branes, 
$\alpha_2$ must be somewhat restricted. Indeed, there are several works on the
 causality violation  of 
 AdS black branes in conjunction with the
 KSS bound~\cite{Brigante:2008gz,deBoer:2009gx,Camanho:2009hu}. 
Clearly, it is important to clarify the pathologies in AdS black branes.

In this paper, we consider the pathologies in AdS black brane
 solutions in Lovelock theory, especially in 10-dimensions and 11-dimensions.
In recent work, this issue has been investigated in \cite{Camanho:2010ru}
based on the master equations derived by us~\cite{Takahashi:2010gz}. 
They used near horizon analysis in the most part of their work
and alluded the importance of the bulk geometry based on the numerical results.
However, they have never given general conditions for the occurrence of pathologies.
In this paper, we explicitly present the conditions for the occurrence of 
pathologies. Using the conditions, we will give a detailed analysis of  
Lovelock AdS black branes and discuss its implications in AdS/CFT.

The organization of this paper is as follows. 
In section~\ref{sec:2}, we review Lovelock theory and
 explain a graphical method of constructing black brane solutions. 
In section~\ref{sec:3}, we clarify the conditions for avoiding the naked singularity,
the ghost instability and the dynamical instability. 
In section~\ref{sec:4}, we examine the pathologies in AdS black branes numerically.
In section~\ref{sec:5}, based on the numerical results, we  discuss
 implications of our findings in the AdS/CFT correspondence. 
The final section~\ref{sec:5} is devoted to the conclusion. 
\section{Lovelock AdS Black Branes}
\label{sec:2}
In this section, we review Lovelock theory and introduce
a graphical method of constructing AdS black brane solutions. 

 The most general  divergence free symmetric tensor constructed out of 
 a metric and its first and second derivatives has been obtained
  by Lovelock~\cite{Lovelock:1971yv}. 
The corresponding Lagrangian can be constructed from $m$-th order Lovelock terms
\begin{eqnarray}
  {\cal L}_m = \frac{1}{2^m} 
  \delta^{\lambda_1 \sigma_1 \cdots \lambda_m \sigma_m}_{\rho_1 \kappa_1 \cdots \rho_m \kappa_m}
  R_{\lambda_1 \sigma_1}{}^{\rho_1 \kappa_1} \cdots  R_{\lambda_m \sigma_m}{}^{\rho_m \kappa_m}
                       \ ,
\end{eqnarray}
where  $R_{\lambda \sigma}{}^{\rho \kappa}$ is the Riemann tensor in $D$-dimensions
and $\delta^{\lambda_1 \sigma_1 \cdots \lambda_m \sigma_m}_{\rho_1 \kappa_1 \cdots \rho_m \kappa_m}$ is the 
generalized totally antisymmetric Kronecker delta defined by 
\begin{eqnarray}
\delta^{\mu_1\mu_2\cdots \mu_p}_{\nu_1\nu_2\cdots\nu_p}={\rm det}
\left(
\begin{array}{cccc}
\delta^{\mu_1}_{\nu_1}&\delta^{\mu_1}_{\nu_2}&\cdots&\delta^{\mu_1}_{\nu_p}\\
\delta^{\mu_2}_{\nu_1}&\delta^{\mu_2}_{\nu_2}&\cdots&\delta^{\mu_2}_{\nu_p}\\
\vdots&\vdots&\ddots&\vdots\\
\delta^{\mu_p}_{\nu_1}&\delta^{\mu_p}_{\nu_2}&\cdots&\delta^{\mu_p}_{\nu_p}
\end{array}
\right)\ .
\nonumber
\end{eqnarray}
By construction, the Lovelock terms vanish for $2m>D$. It is also known that
the Lovelock term with $2m=D$ is a topological term.   
Thus, Lovelock Lagrangian in  $D$-dimensions is defined by
\begin{eqnarray}
  {\cal L} = \sum_{m=0}^{k} c_m {\cal L}_m \ ,   \label{eq:lag}
\end{eqnarray}
where we defined the maximum order $k\equiv [(D-1)/2]$ and  $c_m$ are 
arbitrary constants. 
Here, $[z]$ represents the maximum integer satisfying $[z]\leq z$. 
Taking variation of the Lagrangian with respect to the metric,
 we can derive Lovelock equations
\begin{eqnarray}
	0=\sum_{m=0}^kc_m  \delta^{\nu \lambda_1 \sigma_1 \cdots \lambda_m \sigma_m}_{\mu \rho_1 \kappa_1 \cdots \rho_m \kappa_m}
       R_{\lambda_1 \sigma_1}{}^{\rho_1 \kappa_1} \cdots  R_{\lambda_m \sigma_m}{}^{\rho_m \kappa_m}\ . \label{eq:EOM}
\end{eqnarray}
Hereafter, we set $c_0=(D-1)(D-2)\lambda$ $(\lambda>0)$, $c_1=1$ and $c_m={\alpha}_m/\left\{m\lambda^{m-1}\prod_{p=1}^{2m-2}(D-2-p)\right\}\ (m\geq 2)$ for convenience.  
Note that the coefficients $\alpha_m$ are dimensionless. 

It is well known that there exist static 
exact solutions of the Lovelock equations (\ref{eq:EOM})~\cite{Wheeler:1985nh, Cai:2001dz}. 
Let us consider the following metric
\begin{eqnarray}
   ds^2=r^2\psi(r)dt^2 + \frac{dr^2}{-r^2\psi(r)}+r^2{\delta}_{i j}dx^idx^j 
   \label{eq:solution}\ .
\end{eqnarray}
We assume that $\psi(r)$ is negative outside of the horizon.
Substituting this metric ansatz into Eq.(\ref{eq:EOM}), 
we can obtain an algebraic equation for $\psi(r)$:  
\begin{eqnarray}
      W[\psi]\equiv\sum_{m=2}^{k}\left(\frac{\alpha_m}{m}\psi^m\right)+\psi+1=\frac{\mu}{r^{D-1}}  \ ,
\label{eq:poly}
\end{eqnarray}
where   $\mu$ is a constant of integration which is related to the ADM mass and we assume  
 it is positive. 
Note that we fixed the scale by setting $\lambda=1$ in Eq.(\ref{eq:poly}).

\begin{figure}[htbp]	
\includegraphics[width=7cm]{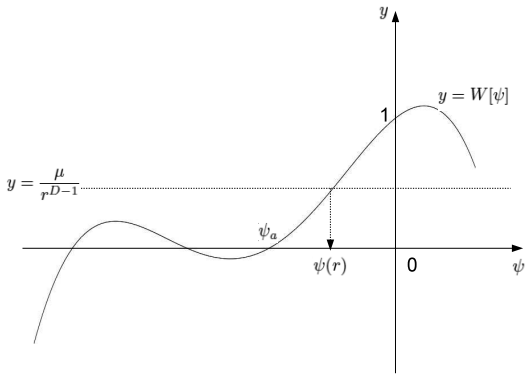}
\caption{A curve $y=W[\psi]$ and a line $y=\mu/r^{D-1}$ are depicted in $\psi-y$ plane. 
Note that $r$ should be regarded as a constant when we draw this figure. 
The intersection of the curve and the line determines a solution of 
Eq.(\ref{eq:poly}). Here, $\psi_a$ is defined as the largest negative root of $W[\psi]=0$. }
\label{fig:1}
\end{figure}

Now, we explain how to construct solutions using a graphical method. 
Apparently, $W$ must be positive. In general, there are many branches. 
In Fig.\ref{fig:1}, 
we depicted $y=W[\psi]$ and $y=\mu/r^{D-1}$ with $r$ fixed in $\psi-y$ plane.   
The intersection of the curve and the line determines $\psi$ once $r$ is given.   
By varying $r$, we obtain the solution of Eq.(\ref{eq:poly}). 
Taking a look at the metric (\ref{eq:solution}), we see that the horizon 
corresponds to $\psi =0$. Hence, a black brane corresponds to the branch 
containing $\psi =0$. 
Next, consider the asymptotic infinity $r\rightarrow \infty$ or 
$y=\mu/r^{D-1}\rightarrow 0$, the function $\psi(r)$ in Fig.\ref{fig:1} 
approaches $\psi_a$ which is defined as the largest negative root of $W[\psi]=0$. 
Thus, the curve between	$\psi =\psi_a$ and $\psi =0$ defines a black brane solution.

\section{Pathologies}
\label{sec:3}

In this section, we list up pathologies in Lovelock AdS black branes.
In particular, we reveal the conditions for the occurrence of pathologies.

\subsection{Naked Singularity}
\begin{figure}[htbp]	
\includegraphics[width=10cm]{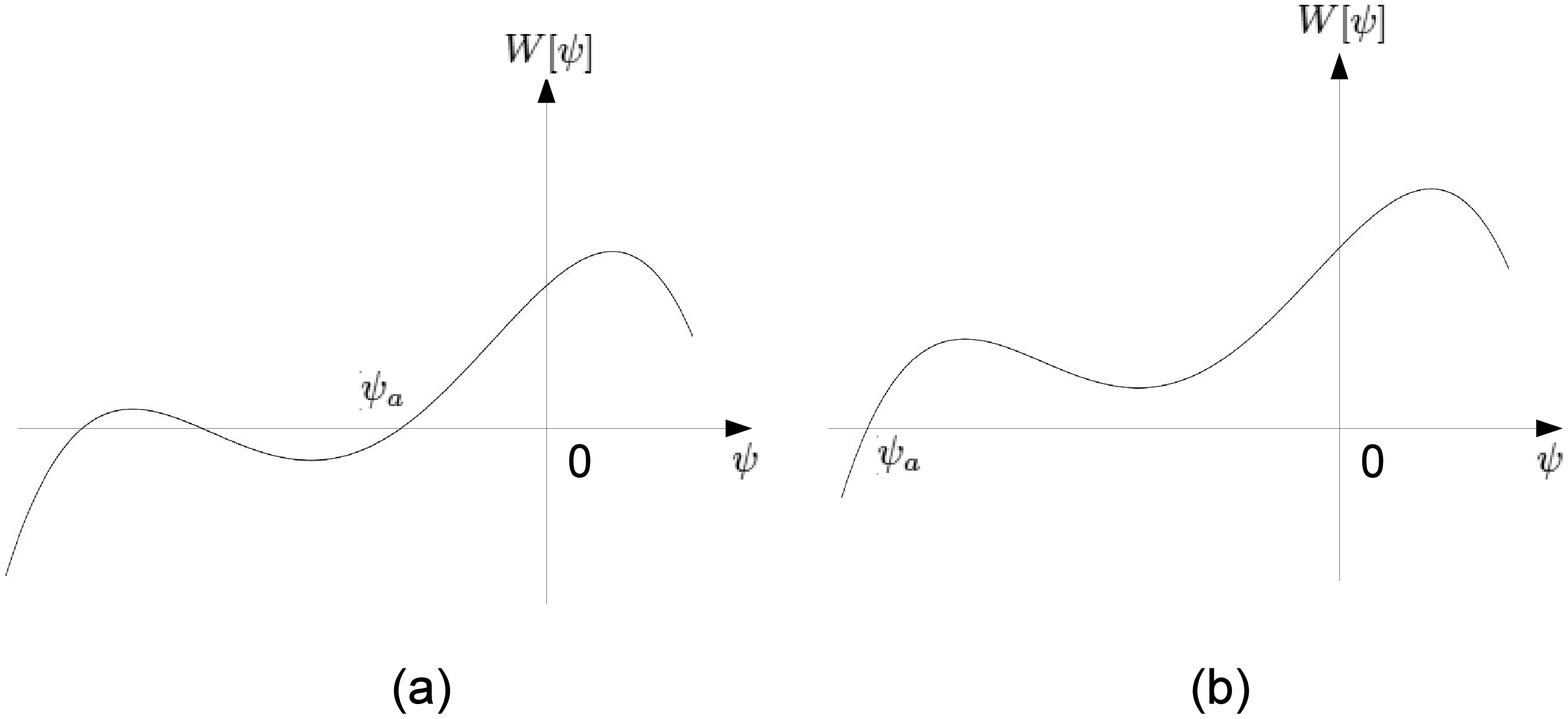}
\caption{(a) Apparently, $W[\psi]$ has no extremal point in the region $[\psi_a,\ 0]$.
Hence, there is no naked singularity in this case.
(b) There is an extremal point. 
 Thus, there exists a naked singularity. }
\label{fig:2}
\end{figure}

In the graphical method, it is easy to find  singularities. 
Let us recall the Kretschmann invariant which is calculated as
\begin{eqnarray}
 R_{\mu\nu\rho\lambda}R^{\mu\nu\rho\lambda} 
 = ( \partial_r^2 (r^2\psi))^2 +2(D-2)\frac{(\partial_r (r^2\psi))^{2}}{r^2}
+2(D-2)(D-3)\psi^2 \ . 
\label{}
\end{eqnarray}
If this invariant diverges, there exist singularities. 
 This occurs at $r=0$ and the point where $\partial_r\psi$ diverges. 
 In fact, $\partial_r\psi$ diverges when $W[\psi]$ becomes extremal value because 
 of a relation  $\partial_r  \psi = -(D-1)W[\psi]/(r\partial_{\psi}W)$
 obtained from (\ref{eq:poly}). 
Since $\partial_{\psi}W|_{\psi=0}=1>0$,  
if $W[\psi]$ is monotonically increasing in the region $[\psi_a,\ 0]$, 
there is no naked singularity. Fig.\ref{fig:2}-(a) corresponds to this case. 
However, like in Fig.\ref{fig:2}-(b), if $W[\psi]$ has an extremal point between $\psi_a$ 
and $0$, there exists a naked singularity. 
Note that the shape of $W[\psi]$ depends only on Lovelock coefficients $\alpha_m$, 
so whether a branch has a naked singularity or not is determined by these constants. 
Since we want to avoid the naked singularity,  we have to exclude the solutions
which have extrema between $\psi = \psi_a$ and $\psi =0$. 
Note that there maybe exotic cases for which $\psi_a$ does not exist.
These solutions should be excluded because they necessarily have the
naked singularity. 

\subsection{Ghost Instability}

In Lovelock theory, the sign in front of the kinetic term in the action
could be negative, namely the ghost instability could occur.
In the previous paper~\cite{Takahashi:2010gz}, we have shown that there exists 
the ghost instability when  
\begin{eqnarray}
K[\psi]  \equiv (D-3)(\partial_{\psi}W)^2-(D-1)W\partial_{\psi}^2 W 
\label{ghost}
\end{eqnarray}
becomes negative. 
Hence, we need to check the sign of $K[\psi]$ to check if a black brane
has the ghost instability or not. 

\subsection{Dynamical Instability}

As we have shown in \cite{Takahashi:2010gz}, the function $W(\psi)$ determines 
if the dynamical instability of Lovelock black branes occurs.
Using the symmetry of the planar part of the metric, we can classify
metric perturbations into the scalar, vector, tensor sectors. 
In the absence of the ghost instability, we can prove that
there is no dynamical instability in the vector sector~\cite{Takahashi:2010gz}. 

There exist the dynamical instability for tensor sector when 
\begin{eqnarray}
L[\psi]&\equiv&
 (D-3)(D-4)\left(\partial_{\psi}W\right)^4
      -(D-1)(D-6) W \left(\partial_{\psi}W\right)^2 \partial_{\psi}^2W \nonumber\\
	&\ &\hspace{1.0cm}+(D-1)^2
      W^2\left\{\partial_\psi W \partial_\psi^3 W-(\partial_\psi^2 W)^2\right\} \ ,
 \label{tensor}     
\end{eqnarray}
is negative~\cite{Takahashi:2009xh,Takahashi:2010gz}. 
Similarly, there exist the dynamical instability for scalar sector, when 
\begin{eqnarray}
M[\psi]&\equiv&
(D-2)(D-3)(\partial_{\psi}W)^4-3(D-2)(D-1)W\partial_{\psi}^2W(\partial_{\psi}W)^2\nonumber\\
&\ &\hspace{1cm} +(D-1)^2W^2\left\{3(\partial_{\psi}^2W)^2-\partial_{\psi}W\partial_{\psi}^3W\right\}.
\label{scalar}
\end{eqnarray}
is negative~\cite{Takahashi:2010gz}. 
In both cases, the square of the effective speed of sound becomes negative.
This kind of instability is found in the cosmological context for the first 
time~\cite{Kawai:1998ab}. 

In order to find the dynamical instability, 
what we have to check is  the sign of $L[\psi]$ and $M[\psi]$ 
in the region $\psi_a<\psi<0$. 
Note that these functions and $\psi_a$ is independent of $\mu$, hence whether 
the dynamical instability exist or not depends only on Lovelock coefficients $\alpha_m$.

\section{Pathology Inspection}
\label{sec:4}

Now, we are in a position to  examine the pathologies in AdS black branes numerically. 
Our strategy  is very simple. 
For each coefficients $\alpha_m$, we search for $\psi_a$ and 
 check the sign of $\partial_{\psi}W$, $K[\psi]$, $L[\psi]$ and $M[\psi]$ in the 
 region $\psi_a<\psi<0$.  The mesh size of this calculation is $\Delta \alpha_m=0.05$. 
In this paper, we concentrate on 10-dimensions and 11-dimensions. 

\subsection{10-dimensions}

\begin{figure}[h]
 \begin{center}
 \begin{tabular}{ccl}
 \includegraphics[width=60mm]{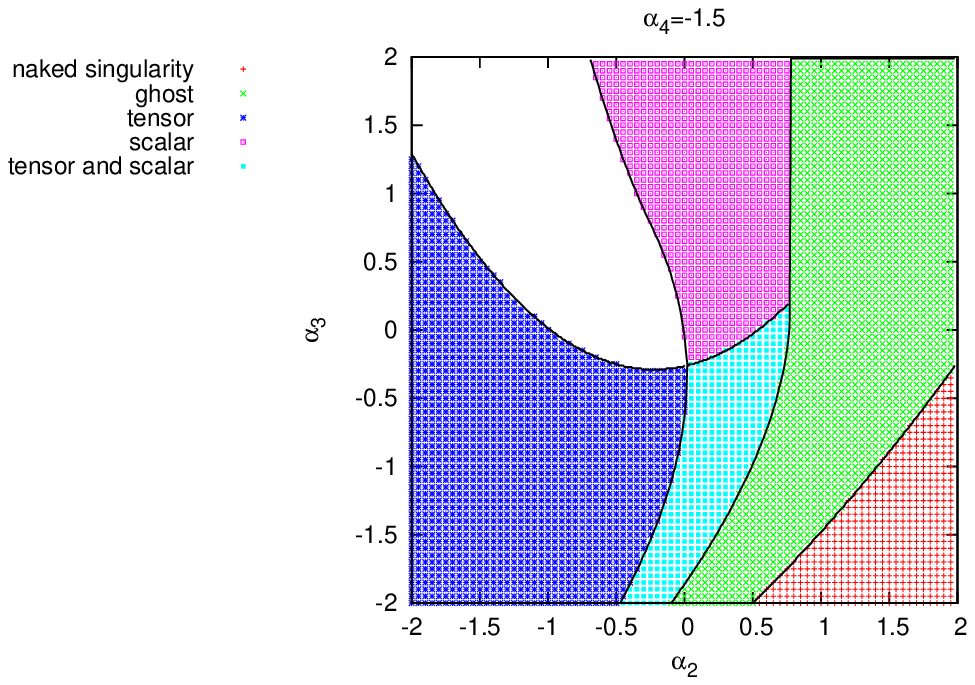}&
\includegraphics[width=60mm]{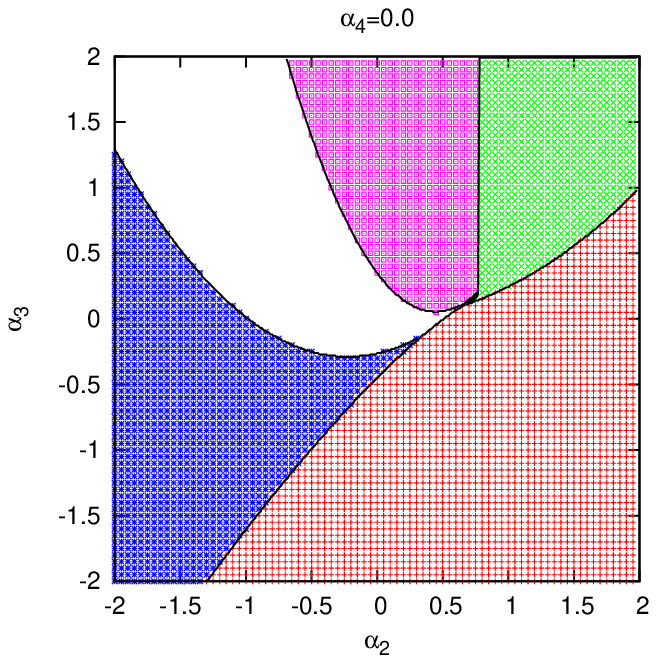}&
\includegraphics[width=60mm]{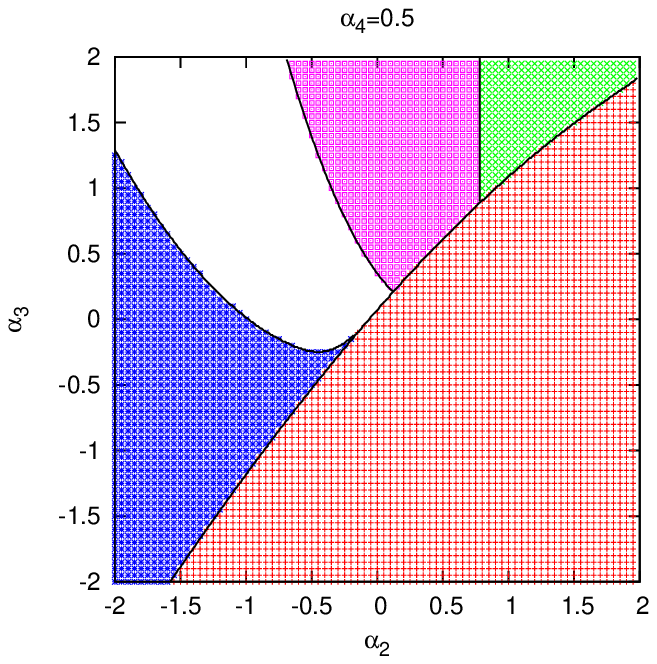}
  \end{tabular}
 \end{center}
 \caption{We plot forbidden regions in $\alpha_2-\alpha_3$ plane with $\alpha_4=-1.5,\ 0,\ 0.5$ respectively. }
 \label{fig:3}
\end{figure}

In 10-dimensions, the Lovelock black holes can be characterized by
the functional
\begin{eqnarray}
  W[\psi] = \frac{\alpha_4}{4} \psi^2 + \frac{\alpha_3}{3} \psi^3 
   + \frac{\alpha_2}{2} \psi^2 + \psi +1 \ .
\end{eqnarray}
Substituting this expression into $\partial_\psi W[\psi]$, (\ref{ghost}),
(\ref{tensor}), and (\ref{scalar}), we can find forbidden region in 
3-dimensional parameter space $\{\alpha_2 , \alpha_3 , \alpha_4 \}$.
In Fig.\ref{fig:3},   
we plot forbidden regions in $\alpha_2-\alpha_3$ plane 
with $\alpha_4=-1.5,\ 0,\ 0.5$, respectively.
The red region is excluded because of the naked singularity, 
the green denotes the models with the ghost instability.
The pink and blue represents the dynamical instability in the scalar and tensor
sector, respectively. In the light blue region, we have the dynamical instability
both in the scalar and the tensor sectors. 

\begin{figure}[h]
 \begin{tabular}{ccl}
  \includegraphics[width=60mm]{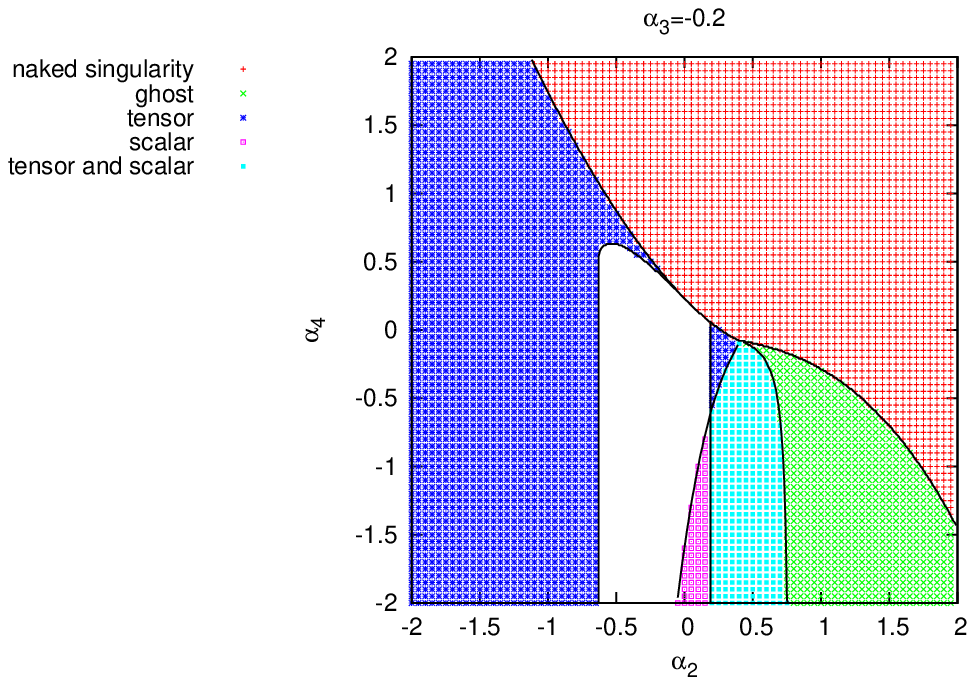}&
  \includegraphics[width=60mm]{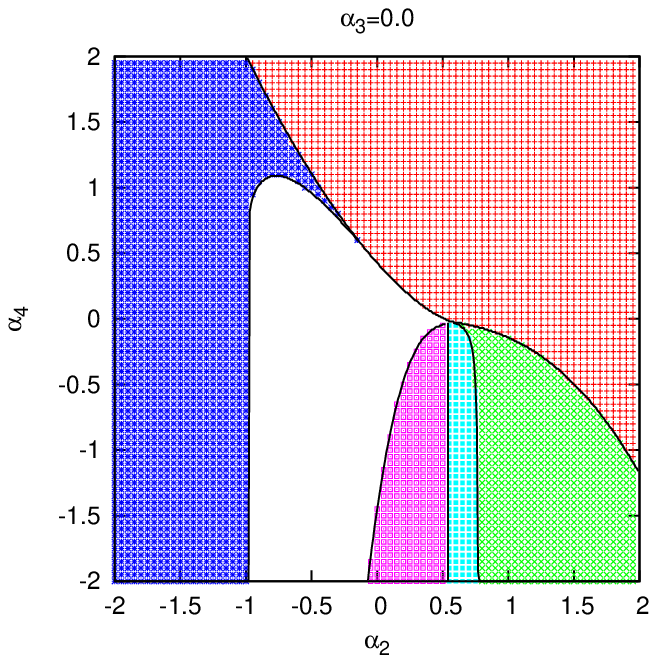}&
  \includegraphics[width=60mm]{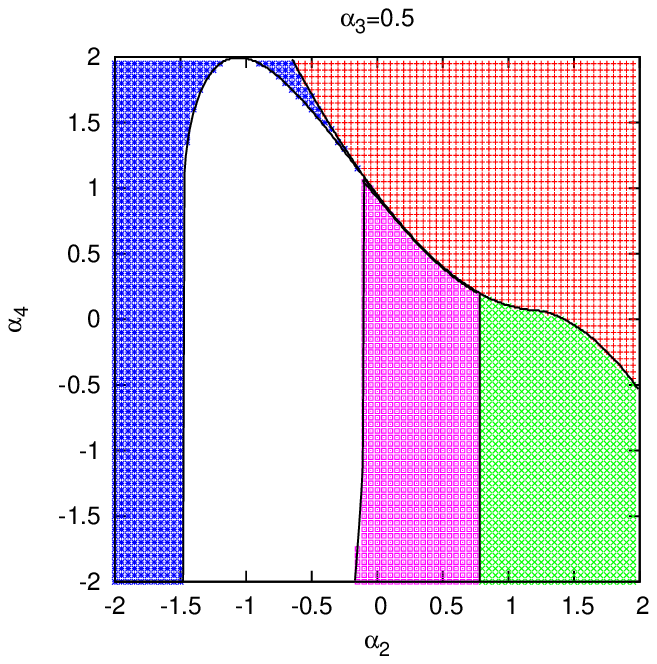}
 \end{tabular}
 \caption{We plot forbidden regions in $\alpha_2-\alpha_4$ plane with $\alpha_3=-0.2,\ 0,\ 0.5$ respectively.}
 \label{fig:4} 
\end{figure}

In Fig.\ref{fig:3}, when $\alpha_4=0$, the border between the white region and the pink region 
can be obtained from the condition $ M[0]=0$ as
\begin{eqnarray}
\alpha_3 =
\frac{3}{2}\alpha_2^2-\frac{3(D-2)}{2(D-1)}\alpha_2+\frac{(D-2)(D-3)}{2(D-1)^2}
\label{}
\end{eqnarray}
Similarly, the border between the white region and the blue region 
can be determined by the condition $L[0]=0$ as 
\begin{eqnarray}
\alpha_3 =
\frac{\alpha^2_2}{2}+\frac{D-6}{2(D-1)}\alpha_2-\frac{(D-3)(D-4)}{2(D-1)^2}\ .
\label{}
\end{eqnarray}
Thus, we see that these dynamical instabilities occur near the horizon 
because $\psi=0$ corresponds to the horizon. 
These results are consistent with those obtained in~\cite{Camanho:2010ru}. 
Note that $M[0]$ and $L[0]$ are determined by $\alpha_2$ and $\alpha_3$ and so these borders are independent of $\alpha_4$ 
if instabilities occur near horizon.
However, comparing three figures in Fig.\ref{fig:3}, 
the region  where black holes are unstable under scalar perturbations 
for $\alpha_4=-1.5$ is very different from that for $\alpha_4=0,\ 0.5$. 
This suggests that these instabilities occur away from the horizon. 
Therefore, $\alpha_4$ affects the behavior of $M[\psi]$  
and change the allowed region in $\alpha_2-\alpha_3$ plane.
Indeed, this fact can be understood more easily from Fig.\ref{fig:4}.  
In Fig.\ref{fig:4}, 
we plot forbidden regions in $\alpha_2-\alpha_4$ plane with $\alpha_3=-0.2,\ 0,\ 0.5$, respectively.
In these figures, we see vertical stripes for negative $\alpha_4$. 
For example, in Fig.\ref{fig:4} with $\alpha_3=0$, 
there are three vertical lines: 
$\alpha_2\simeq -1,\ \alpha_2\simeq 0.5, \ \alpha_2\simeq 0.75$. 
These lines can be obtained from $L[0] =0$ as
\begin{eqnarray}
\alpha_2 = -\frac{D-6}{2(D-1)}\pm \sqrt{2\alpha_3+\frac{5D^2-40D+84}{4(D-1)^2}} 
\label{}
\end{eqnarray}
and from $K[0]=0$ as 
\begin{eqnarray}
 \alpha_2 = \frac{D-3}{D-1} \ .
\label{}
\end{eqnarray}
However, when $\alpha_4$ becomes large, the stripe structure collapses. 
This suggests instabilities are not originated from the near horizon geometry. 
It turned out that $\alpha_4$ is a relevant parameter for AdS black branes. 

\subsection{11-dimensions}

\begin{figure}[h]
 \begin{tabular}{ccl}
  \includegraphics[width=60mm]{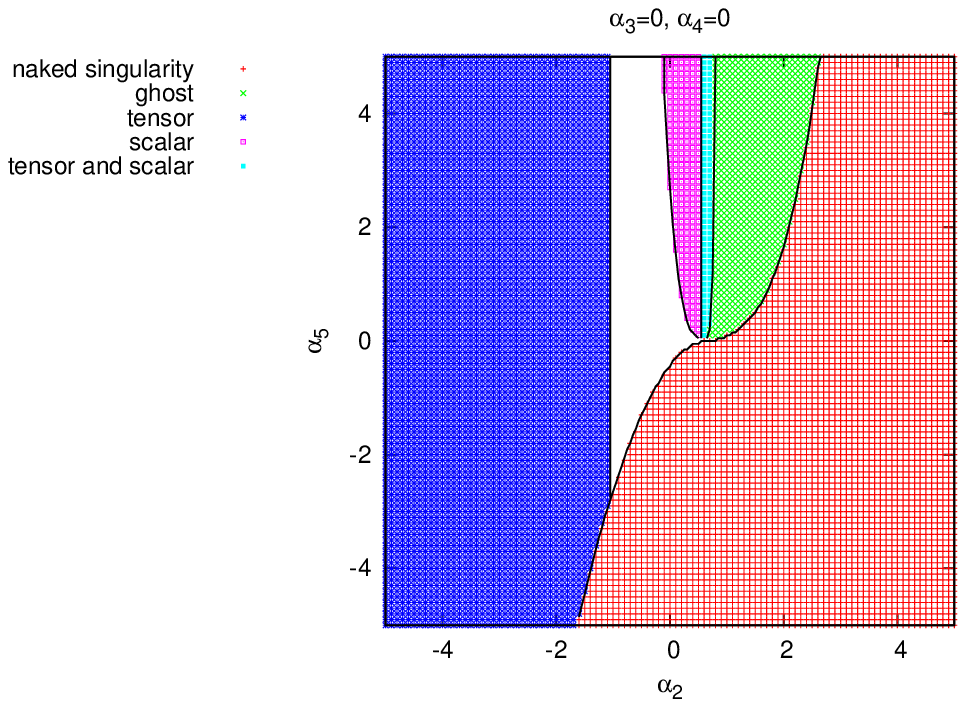}&
  \includegraphics[width=60mm]{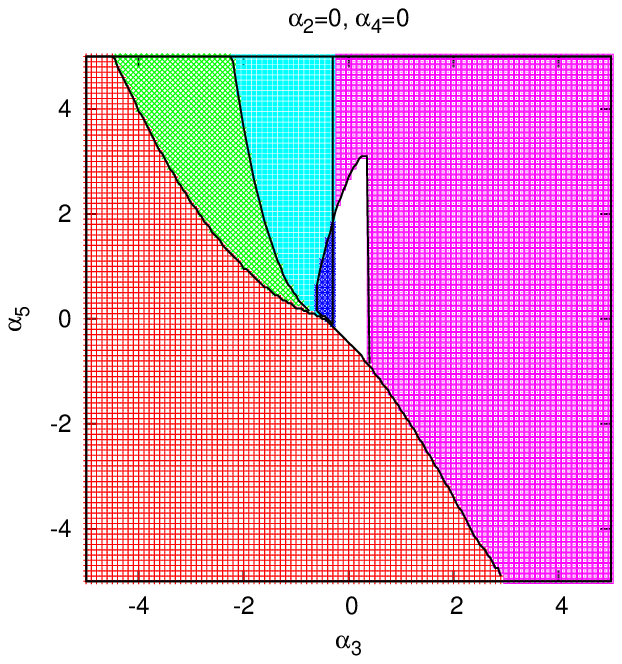}&
  \includegraphics[width=60mm]{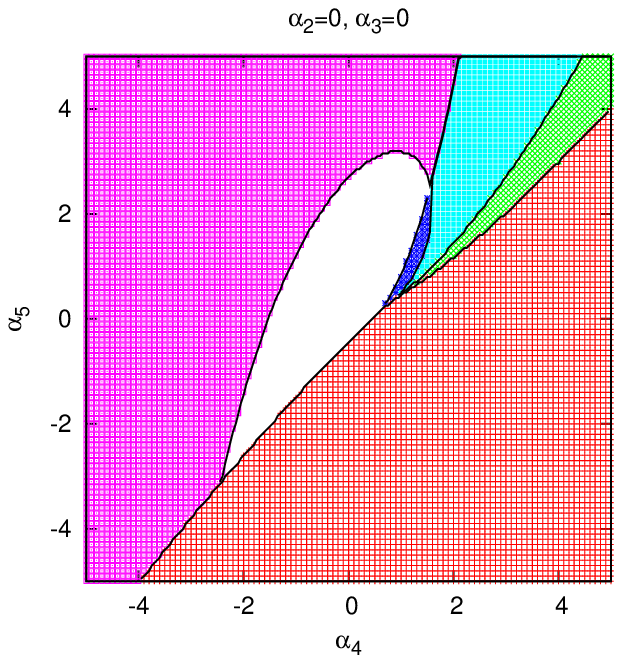}
 \end{tabular}
 \caption{We plot forbidden regions in $\alpha_2-\alpha_5$, $\alpha_3-\alpha_5$
  and $\alpha_4-\alpha_5$ planes, respectively.
     In these figures, we set other Lovelock coefficients to be $0$.}
 \label{fig:5} 
\end{figure}

In 11-dimensions, the key functional is given by
\begin{eqnarray}
  W[\psi] = \frac{\alpha_5}{5} \psi^2
  + \frac{\alpha_4}{4} \psi^2 + \frac{\alpha_3}{3} \psi^3 
   + \frac{\alpha_2}{2} \psi^2 + \psi +1 \ .
\end{eqnarray}
Again, substituting this expression into $\partial_\psi W[\psi]$, (\ref{ghost}),
(\ref{tensor}), and (\ref{scalar}), we can find forbidden region in 
4-dimensional parameter space $\{\alpha_2 , \alpha_3 , \alpha_4 ,\alpha_5 \}$.
Of course, it is a formidable task to visualize such a higher dimensional space.
Hence, we look at several sections in the parameter space.
In Fig. \ref{fig:5}, we plot forbidden regions  in $\alpha_2-\alpha_5$ plane with $(\alpha_3,\ \alpha_4)=(0,\ 0)$, 
$\alpha_3-\alpha_5$ plane with $(\alpha_2,\ \alpha_4)=(0,\ 0)$ and $\alpha_4-\alpha_5$ plane with $(\alpha_2,\ \alpha_3)=(0,\ 0)$, respectively. 

From these figures, we see that $\alpha_5$ affects the allowed ranges 
of $\alpha_2$, $\alpha_3$ and $\alpha_4$. In particular, in the case of 
$\alpha_3-\alpha_5$ and $\alpha_4-\alpha_5$ planes, the allowed region 
is finite. It indicates that AdS black branes in Lovelock theory
are pathological in most cases. 


\section{Implications in AdS/CFT}
\label{sec:5}

Let us discuss implications of our numerical results in
the AdS/CFT correspondence. 

With the master equation in \cite{Takahashi:2009xh},
the shear viscosity to entropy ratio  $\eta/s$ has been calculated as 
\begin{eqnarray}
\frac{\eta}{s}=\frac{1}{4\pi}\left(1-\frac{D-1}{D-3}\alpha_2\right)\nonumber
\label{}
\end{eqnarray}
through AdS/CFT correspondence~\cite{Shu:2009ax}. 
Note that this depends only on $\alpha_2$. 
Hence, it seems that $\alpha_3$, $\alpha_4$ and $\alpha_5$ do not affect this value. 
However, as our numerical calculations have shown, 
$\alpha_3$, $\alpha_4$ and $\alpha_5$ affect the allowed region of $\alpha_2$. 
This fact was also noticed in \cite{Camanho:2010ru}.
Interestingly, from Fig.\ref{fig:3}, we see that a
positive $\alpha_2$ is not allowed for any $\alpha_3$ if $\alpha_4=-1.5$. 
This means that the bound of $\eta/s$ must be larger than $1/4\pi$ if $\alpha_4=-1.5$. 
Thus, it turned out that the KSS like bound is sensitive to Lovelock coefficients.

It is also possible to apply our results to holographic 
superconductors~\cite{Hartnoll:2008vx}. 
There, the universality for the ratio between the frequency dependent conductivity
and the critical temperature is found~\cite{Horowitz:2008bn}.
In Gauss-Bonnet theory, it is pointed out that
this universality in holographic superconductors
is violated for a large $\alpha_2$~\cite{Gregory:2009fj}. 
However, it is probable that this violation is due to 
the pathologies discussed in this paper. It would be interesting to
extend holographic superconductors to Lovelock theory to clarify
this point.

\section{Conclusion}
\label{sec:6}

We have discussed the pathologies in AdS black branes 
in Lovelock theory, especially in 10-dimensions and 11-dimensions.
We obtained the general conditions for Lovelock coefficients $\alpha_m$
 that these black branes have the naked singularity,
the ghost instability, and the dynamical instability.
It turned out that the dynamical instability could occur away from the horizon
in contrast to a naive expectation.  
Thus, $\alpha_4$ and $\alpha_5$ also control the allowed region of $\alpha_2$, and consequently changes the lower limit of $\eta/s$.  
We have also pointed out that the pathologies we have found could affect
the interpretation of higher dimensional holographic superconductors.

In this paper, we did not consider the causality violation discussed in \cite{Brigante:2008gz,deBoer:2009gx,Camanho:2009hu,Camanho:2010ru}. It is easy to take into account the 
causality violation based on the master equations~\cite{Takahashi:2010gz}. Then,
we could further restrict the allowed region for $\alpha_2$. 
It is also straightforward to extend our analysis to other dimensions
using the master equations~\cite{Takahashi:2010gz}. 

\begin{acknowledgements}
We wish to thank Keiju Murata for fruitful discussions. 
This work was supported in part by the
Grant-in-Aid for  Scientific Research Fund of the Ministry of 
Education, Science and Culture of Japan No.22540274, the Grant-in-Aid
for Scientific Research (A) (No.21244033, No.22244030), the
Grant-in-Aid for  Scientific Research on Innovative Area No.21111006,
JSPS under the Japan-Russia Research Cooperative Program,
the Grant-in-Aid for the Global COE Program 
``The Next Generation of Physics, Spun from Universality and Emergence".
\end{acknowledgements}

\end{document}